 \documentclass[aps,prc,showpacs,floatfix,4paper]{revtex4}
\usepackage{amsmath}
\usepackage{amssymb}
\usepackage{epsfig}
\usepackage{graphicx}
\def\bfg #1{{\mbox{\boldmath $#1$}}}

\begin{document}
\title{Null-test signal for T-invariance violation in $pd$ scattering}
%Antiproton scattering off $^3{\rm He}$ and $^4{\rm He}$ nuclei
% at low and intermediate energies}

\author {Yu.N. Uzikov$^{a,b}$, A.A. Temerbayev$^{c}$}

\affiliation{
$^a$Laboratory of Nuclear Problems, Joint Institute for Nuclear
Research,  Dubna, 141980 Russia\\
$^b$ Department of Physics, Moscow State University,  Moscow, 119991  Russia\\
$^c$L.N. Gumilyov Eurasian National University, Astana, 010008 Kazakhstan
}

\begin{abstract}
  The integrated
 proton-deuteron cross section $\widetilde\sigma$  for the
 case  of the incident proton  vector polarization $p_y^p$ and tensor
 polarization $P_{xz}$ of the deuteron target provides a null test signal
 for  time-reversal invariance violating but P-parity conserving (TVPC)
 effects.
 We study the null-test observable
 $\widetilde\sigma$   within the Glauber theory of the
 double-polarized $pd$ scattering.
 Full spin dependence of the ordinary strong $pN$ scattering amplitudes
 and  different types of the hypothetical TVPC pN-amplitudes
%interaction
 are taken into account.
 We show that the contribution from the exchange of the lowest-mass meson
% allowed   the  $\rho$-meson exchange, which is
% the lowest mass-meson
  allowed in the TVPC interaction, i.e. the $\rho$-meson,
  to the null-test observable
 $\widetilde\sigma$ is zero. The axial $h_1$ meson exchange makes a non-zero
 contribution.
 We find that inclusion of the Coulomb
 interaction does not lead to divergence of the cross section
 $\widetilde\sigma$ and calculate its energy  dependence at the proton beam energy
 100-1000 MeV.

%
%%%%%%%%%%%%%%%%%%%%%%%%%%%%%%%%%%%%%555
%

\end{abstract}


\maketitle

\section{Introduction}
% CP-violation observed in physics of kaons \cite{kronin64} and B-mesons \cite{Bcp}
% suggests T-invariance violation  assuming  that the CPT -invariance is valid.
CP violation (or T-reversal invariance violation under CPT symmetry) is required to explain the
 baryon asymmetry of the Universe \cite{Sakharov}. In baryon systems  violation of  T-invariance
 has not been observed yet.
 CP violation  established
  in physics of kaons
  % \cite{kronin64}
   and B-mesons
% \cite{Bcp}
  leads to simultaneous  CP and P-invariance violation.
  Under the assumption of CPT-invariance
 this implies existence of T-odd  P-odd interactions. These effects are parametrized  in
  the Standard Model by the
 CP violating phase of the Cabibbo-Kobayashi-Maskawa matrix. Another source for T-odd P-odd
 effects is the QCD  $\theta$-term, which can be related to electric dipole moments  (EDM)
  of elementary particles and atoms in their ground states.

  On the contrary,
% time-reversal-non-invariant (T-odd)
time-reversal symmetry violating  (T-odd) P-parity conserving
 (P-even) flavor conserving (TVPC) interactions do not arise on the fundamental level within
the Standard Model,
%  in first order of the boson exchanges
% between fermions.
 although they can be generated from the T-odd P-odd interaction by
 weak radiative P-parity non-conserving corrections.
% For the first time this kind of interaction was  supposed by Okun \cite{okunj}
% in order to explain the CP-violation in the kaons decays with assumed intensity $\alpha_T\sim 10^{-3}$.
% Later on
However in this  case its intensity  is too low \cite{Khripl91,Gudkov92}
to be observed in experiments at present. Thus, observation of  the TVPC effects
would be considered as indication of physics beyond the standard model.

 The existing  experimental constraints on the TVPC effects in
  physics of nuclei are rather weak. So, the test of the detailed balance performed
   for the reactions
   $^{27}Al(p,\alpha)^{24}Mg$ and  $^{24}Mg(\alpha,p)^{27}Al$
  \cite{blanke},
 and  complemented by numerous statistical analyses of
 nuclear energy-level fluctuations  leads to the ratio of T-odd to T-even matrix elements as
 $\alpha_T < 2\times 10^{-3}$ \cite{french}.
Another type of experiment, i.e. polarized neutron transmission through a polarized $^{165}$Ho target
gives $ \alpha_T \leq 7.1 \times 10^{-4}$
or $\bar g_\rho \leq 5.9\times 10^{-2}$ \cite{huffman}.
Here $\bar g_\rho$ is the T-odd P-even coupling constant of the charged $\rho$-meson with the nucleon
  introduced in Ref. \cite{simonius97} to classify the TVPC interactions
 in terms of boson exchanges.
 Charge symmetry breaking  determined as difference
%CSB $\Delta A$ ($A=A^n-A^p$)
 in scattering of  polarized protons off unpolarized neutrons ${\vec p}n$ and polarized neutrons off
 unpolarized  protons
 ${\vec n}p$ gives
 $\alpha_T \leq 8 \times 10^{-5}$ (or $\bar g_\rho< 6.7\times 10^{-3})$ \cite{simonius97}.
  One should add that  indirect model-dependent estimation  based on the existing
 constraints on  EDM  gives
  $ \alpha_T \leq 1.1 \times 10^{-5}$ ($\bar g_\rho\leq \times 10^{-3}$) \cite{haxtonHM94}.
However, a more recent analysis showed \cite{kurylov} that EDM may arise  via another
 scenario which  suggests no significant  constraints on the TVPC forces.

%%%%%%%%%%%%%%%%%%%%%%%%%% from FEw-Body-21-chikago
 The integrated cross section $\widetilde\sigma$
 will be measured   at COSY \cite{TRIC} in double polarized  $pd$ scattering with a
 transverse polarized proton beam ($p_y^p$) and a tensor polarized deuterium
 target ($P_{xz}$). This observable provides a real null test of the TVPC forces \cite{conzett}.
 This signal
 is not affected by the initial and final state interaction and therefore its observation
 would directly indicate time-invariance violation,  as in case of the neutron EDM.
 The  experiment \cite{TRIC} will be performed  at  a beam energy of 135 MeV. This energy choice
  was motivated by the theoretical   analysis
 of the integrated $pd$
  cross section ${\widetilde \sigma}$
 performed in Ref.  \cite{beyer}.
  The aim of this experiment is to diminish  the upper bound on the TVPC effects previously obtained
%results of previous
    in the  ${\vec n}^{167}$Ho scattering \cite{huffman} by one order of magnitude.

%%CHECK 
 The elastic channel and the deuteron breakup $dp\to pnp$  were considered in
  Ref. \cite{beyer} in the impulse approximation (single scattering mechanism)  
 for estimation of ${\widetilde\sigma}$.
  In the present work we  study the null-test observable
${\widetilde \sigma}$ on the basis of the generalized optical theorem using the forward elastic
 $pd$ scattering amplitude calculated within  the   Glauber theory.
 Both  the single and double scattering mechanisms are 
 considered.
 %%%%%% END of CHECK
 The spin-dependent  Glauber  formalism  recently developed in Ref. \cite{PK} was
 applied in our previous work
 \cite{TUZyaf} to calculate spin observables of the elastic
 $pd$ scattering   using  the
 strong (time invariance  conserving and P-parity conserving)
 $pN$ scattering  amplitudes as input at 135 MeV. The obtained
  differential cross section,
 vector and tensor analyzing powers and spin-correlation parameters were found to be in
reasonable agreement  with the existing data \cite{sekiguchi,przewoski}.
 Here we   generalize this formalism
 to allow for TVPC $pN$ scattering amplitudes of several types. This  generalized  formalism is
 applied below to derive formulas for the null-test observable
${\widetilde \sigma}$ and calculate its energy dependence. We show that within the single
 scattering mechanism this observable is zero in the Glauber theory (for any type of the TVPC
 $pN$  interactions considered in the general case in Ref. \cite{herczeg})   and,
 consequently, focus on the double scattering mechanism. We investigate the
 contribution of several
 TVPC terms to the $pN$ scattering amplitudes, in particular, the $\rho$-meson  and axial $h_1$- meson
 exchanges. In addition, we investigate the influence of the Coulomb interaction on
 the ${\widetilde \sigma}$ cross section not considered in Ref.\cite{beyer}.

 The paper is organized as follows. In Sect. II we consider the spin structure of the forward
 $pd$ elastic scattering amplitude including the TVPC term and apply the generalized optical
 theorem to derive formulas for total spin-dependent cross sections in terms of the forward
 scattering invariant amplitudes. In Sect. III we  construct the Glauber scattering operator
 taking into account full spin dependence of the elementary $pN$-scattering amplitudes
 for strong and some types of TVPC interactions and S- and D- components of the deuteron
 wave function.
 Analytical expressions for
 the TVPC forward scattering amplitude
 ${\widetilde g}$ are  derived for the double scattering mechanism  with  different
 TVPC terms.
%$g'$, $g$ and $h$ terms.
 The influence of the Coulomb effects on the
 ${\widetilde g}$ amplitude is discussed in Sect.IV.  Numerical  results are shown in Sect.IV.

\section{Forward transition operator and integrated cross sections}
Time-reversal symmetry conserving and P-parity conserving (TCPC or T-even P-even)
  interactions lead to the following transition amplitude
  of the elastic $pd$ scattering at zero degree \cite{rekalo}
\begin{eqnarray}
{e_\beta^\prime}^* M(0)^{TCPC}_{\alpha\beta} e_\alpha=
{g_1}[{\bf e\, e}{^\prime}^*-({\bf { m}\bf e})({\bf { m} } {\bf {e^\prime}}^*)]
 + {g_2}(\bf { m} \bf e)(\bf {  m} {\bf {e^\prime}}^*)+\nonumber \\
i {g_3} \{{\bfg  \sigma} [{\bf e}\,\times  {\bf e}{^\prime}^*] -
({\bfg \sigma} {\bf {  m}})({\bf{ m}}\cdot
[{\bf e}\,\times { \bf e^\prime}^*])\}+
i {g_4} ({\bfg \sigma} \bf{  m}) ( \bf{  m} \cdot
 [{\bf e}\,\times  {e^\prime}^*]),\mathrm{}
\label{fab}
\end{eqnarray}
where ${\bf e}$ ($\bf e^\prime$) is the  polarization vector of the initial (final) deuteron,
${\bf { m} }$ is the unit vector along the beam momentum, ${\bfg  \sigma}$ is the Pauli matrix,
 $g_i$ ($i=1,\dots,4)$ are  complex amplitudes.
 To the right-hand side of Eq.(\ref{fab}) one  can add the TVPC (T-odd P-even)
term in a very general form
\begin{eqnarray}
{e_\beta^\prime}^* M(0)^{TVPC}_{\alpha\beta} e_\alpha=
{\widetilde g} \{({\bfg\sigma} \cdot [\bf{  m}\times {\bf e}])({\bf m}\cdot {{\bf e}^\prime}^*)
+({\bfg\sigma} \cdot [\bf{ m}\times {{\bf e}^\prime}^*])({\bf m}\cdot{\bf e} )
\},
\label{fabtvpc}
\end{eqnarray}
where  $\widetilde {g}$  is the TVPC transition amplitude.
 To find  the total spin dependent $p d$ cross sections
 we use the generalized optical theorem
% that has the following form
 \cite{phillips}
% According to \cite{phillips} one has
\begin{equation}
\label{optth}
%Im \frac {Tr(\hat \rho_i \hat F(0))}{Tr \hat \rho_i}=
%\frac{k_{pd}}{4\pi} \sigma_i^t,
\sigma_i^t={4\sqrt{\pi}}Im \frac {Tr( \rho_i  M(0))}{Tr \hat \rho_i},
 \end{equation}
 where
$ M(0)=M(0)^{TCPC}+M(0)^{TVPC}$ is the transition operator from Eqs. (\ref{fab}) and
 (\ref{fabtvpc}) for the $ p d$ elastic
 scattering at  zero  angle $\theta=0$,
 $ \rho_i$ is the initial
 spin-density matrix, $\sigma_i^t$ is the total cross section corresponding to
 the density matrix $\rho_i$. The transition operator  is normalized according to
 the following relation with the differential cross section \cite{PK}
\begin{eqnarray}
\label{normapd}
\frac{d\sigma}{dt}= \frac{1}{6}Tr MM^+.
\end{eqnarray}
 For the sum of Eqs. (\ref{fab}) and (\ref{fabtvpc})
 one can write  (see also \cite{uzpepan98})
\begin{equation}
\label{fabechay}
 M(0)_{\alpha\beta}=g_1\delta_{\alpha\beta}+
(g_2-g_1)m_\alpha k_\beta+ig_3{ \sigma}_i
\epsilon_{\alpha\beta  i}+ i(g_4-g_3){\ \sigma}_i m_i m_j
\epsilon_{\alpha\beta  j}
+{\widetilde g} \sigma_i(\epsilon_{z\alpha i}{ m}_\beta{ m}_z+
\epsilon_{z\beta i}{ m}_z{ m}_\alpha),
\end{equation}
where $ \sigma_i$ ($i=x,y,z$) are the Pauli spin matrices,
$\epsilon_{\alpha \beta \gamma}$ is the fully antisymmetric tensor,
$m_\alpha$  ($\alpha=x,y,z$) are the Cartesian components of the  vector ${ {\bf m}}$.
%directed along the beam momentum.
%, $g_i$ $(i=1,\dots,4)$ are complex numbers
%determined by the dynamics of the reaction. Let us put  the OZ-axis along the
%vector ${\bf m}$, so ${\bf m}=(m_x,m_y,m_z)=(0,0,1)$.

%%%%%%%%%%
 The initial state spin density matrix $\rho_i=\rho^{p}\rho^{d}$ is
 the product of the spin density matrices for the proton
\begin{equation}
\label{rhop}
\rho^{p}=\frac{1}{2}(1+{\bf  p}^{ p}{ {\bfg \sigma}}),
\end{equation}
 where ${\bf  p}^{ p}$ is the polarization vector of the proton, and for the deuteron
\begin{equation}
\label{rhod}
\rho^{d}=\frac{1}{3}+ \frac{1}{2}S_jp_j^d+\frac{1}{9} S_{jk}P_{jk};
\end{equation}
 here $S_j$ is the spin-1 operator,
 ${ p}_j^d$ and $P_{jk}$ $(j,k=x,y,z)$ are the vector
 and tensor polarizations of the deuteron, and
 $S_{jk}=(S_jS_k+S_kS_j-\frac{4}{3}\delta_{jk})$ is the spin-tensor operator.
 Using Eqs.
% (\ref{fab}), (\ref{fabtvpc}),
(\ref{optth}), (\ref{fabechay}),  (\ref{rhop}) and  (\ref{rhod}),
 one can find
  the total cross section of the $pd$ scattering
%(for  normalization of the amplitudes according to Ref.\cite{PK}!!)
 as
 \begin{equation}
\label{totalspin}
{
\sigma_{tot}= {\sigma_0^t+\sigma_1^t{{\bf p}^{ p}\cdot {\bf p}^d}+
 \sigma_2^t {({\bf p}^{ p}\cdot {{\bf m}}) ({\bf p}^d\cdot { {\bf m}})}+
\sigma_3^t { P_{zz}}} +{\widetilde \sigma} {p_y^p P_{xz}^d},
}
\end{equation}
where ${\bf p}^p$  (${\bf p}^d$) is the  vector polarization of the  initial proton (deuteron) and
$P_{zz}$ and $P_{xz}$ are the tensor polarizations of the deuteron. The OZ axis is directed along
% the proton beam momentum
the ${{\bf m}}$, the OY axis is directed along the vector polarization of
 the  proton beam ${\bf p}^p$
 and the OX axis is chosen to form the right-hand reference frame.
 The following equations were found  in \cite{UJH13}
 for the  TVPC
 %T-even P-even
 terms
 \begin{eqnarray}
\label{sigmai}
\sigma_0^t=\frac{4}{3}\sqrt{\pi} Im(2g_1+g_2), \ \ \ \ \sigma_1^t=-4\sqrt{\pi}Im g_3, \\ \nonumber
 \sigma_2^t=-4\sqrt{\pi}Im(g_4- g_3), \ \ \ \  \sigma_3^t=4\sqrt{\pi}Im (g_1-g_2).
\end{eqnarray}
We find that the TVPC term $\widetilde g$ in the forward $pd$ elastic scattering amplitude (\ref{fabtvpc})
  leads
 to the following
 % result for the corresponding
 integrated cross section
\begin{eqnarray}
\label{sigma5}
{\widetilde\sigma}=-4\sqrt{\pi}Im\frac{2}{3}{\widetilde g}.
\end{eqnarray}

 In Eq. (\ref{totalspin}) the terms $\sigma_i^t$ with $i=0,1,2,3$ are non-zero only
 for the TCPC (T-even P-even) interactions and
 the last term ${\widetilde \sigma}$ is non-zero if the TVPC
   (T-odd P-even) interactions  occur.
 Thus, the  term ${\widetilde \sigma}$ constitutes the null-test signal for time-reversal invariance violating but
  P-parity conserving effects.
 This term can be measured in the transmission experiment \cite{TRIC} as  a
  difference of counting
 rates for the cases with
 $p_y^p P_{xz}^d>0$ and  $p_y^p P_{xz}^d<0$.

We find  the following  matrix elements of the TVPC transition operator
(\ref{fabtvpc}):
\begin{eqnarray}
\label{me}
<\mu'=\frac{1}{2},\lambda'=0|M^{TVPC}|\mu=-\frac{1}{2},\lambda=1>
 =i\sqrt{2}{\widetilde g}, \\
 %\nonumber \\
 \label{me2}
 <\mu'=\frac{1}{2},\lambda'=-1|M^{TVPC}|\mu=-\frac{1}{2},\lambda=0> =-i\sqrt{2} {\widetilde g},
 %\nonumber \\
 \label{metp1}
% <\mu'=\frac{1}{2},\lambda'=0|M^{TPC}|\mu=-\frac{1}{2},\lambda=1> =\sqrt{2}g_3, \\
 %\nonumber \\
 \label{metp2}
% <\mu'=\frac{1}{2},\lambda'=-1|M^{TPC}|\mu=-\frac{1}{2},\lambda=0> =\sqrt{2}g_3.
\end{eqnarray}
where $\mu$ ($\mu'$) and $\lambda$ ($\lambda'$) are spin projections of the initial (final)
 proton and deuteron on the beam direction, respectively.
The diagonal matrix elements of the $M^{TVPC}$ operator are zeros.
For the operator $M^{TCPC}$ the corresponding matrix elements in Eqs.(\ref{me}) and (\ref{me2}) are identical 
and  equal to $\sqrt{2}g_3$.
%symmetric with exchange of initial and 
%final states:
%\nonumber \\
%\begin{eqnarray}
%\label{metcpc}
% \label{metp1}
% <\mu'=\frac{1}{2},\lambda'=0|M^{TPC}|\mu=-\frac{1}{2},\lambda=1> =\sqrt{2}g_3, \\ \nonumber \\
% \label{metp2}
% <\mu'=\frac{1}{2},\lambda'=-1|M^{TPC}|\mu=-\frac{1}{2},\lambda=0> =\sqrt{2}g_3.
%\end{eqnarray}

\section{Spin-dependent Glauber formalism with strong and TVPC interaction}

\subsection{Hadronic and Coulomb $pN$ interaction}
% and OY axis is determined by the vector polarization of the proton beam.
Hadronic amplitudes of $pN$ scattering are taken in a form of \cite{PK}
\begin{eqnarray}
\label{pnamp}
M_N({\bf p}, {\bf q};\bfg \sigma, {\bfg \sigma}_N)=
 A_N+C_N\bfg \sigma \hat{\bf  n} +C_N^\prime\bfg \sigma_N \hat{\bf  n }+
B_N(\bfg \sigma \hat {\bf k}) (\bfg \sigma_N \hat {\bf k})+\\ \nonumber
+ (G_N+H_N)(\bfg \sigma \hat {\bf q}) (\bfg \sigma_N \hat {\bf q})
+(G_N-H_N)(\bfg \sigma \hat {\bf n}) (\bfg \sigma_N \hat {\bf n}),
\end{eqnarray}
where  ${\hat {\bf q}}$, ${\hat {\bf k}}$ and ${\hat {\bf n}}$
 are defined as unit vectors along the vectors  ${ {\bf q}}=({\bf p}-{\bf p}')$,
%/|{\bf p}-{\bf p}'|$,
${ {\bf k}}=({\bf p}+{\bf p}')$
%/|{\bf p}+{\bf p}'|$,
and ${ {\bf n}}=[ {\bf k}\times {\bf q}]$,
%/|[{\hat {\bf k}}\times {\hat {\bf q}}']|$
respectively. Normalization of the amplitudes $A_N$, $B_N$, $C_N$, $C^\prime_N$, $G_N$, $H_N$
is the same as in Ref. \cite{PK}
\begin{eqnarray}
\frac{d\sigma}{dt}=\frac{1}{4}Tr M_{N}M_{N}^+,
\label{normapN}
\end{eqnarray}
where $d\sigma/dt$ is the differential cross section
 of the elastic $pN$ scattering. The
 Glauber formalism for the $pd$ elastic scattering
accounting  full spin dependence
of the $pN$ amplitudes (\ref{pnamp}) and S-and D-components of the deuteron wave function is given
in Ref.\cite{PK}. The unpolarized differential cross section and analyzing powers of $pd$
 scattering calculated
 within this formalism are in reasonable agreement with existing data in the forward hemisphere at
 energies 250-1000 MeV \cite{PK}, \cite{jhuz2012}.
 Further development of this formalism  was  done in \cite{TUZyaf} to allow for
 calculation of spin correlation
 parameters and inclusion of the Coulomb interaction that
 is important at lower energies.

 We include  the Coulomb interaction by adding to the Glauber hadronic $pd$ scattering
amplitude the following  pure Coulomb amplitude of
the $pd$ scattering
\begin{eqnarray}
\label{kulonpd}
 %M_{pd}^C({\bf q})=\frac{k_{pd}}{k_{pp}}f_{pp}^C({\bf q}) S_d({\bf q}/2)\times \frac{\sqrt{\pi}}{k_{pd}}.
 M_{pd}^\textrm C({\bf q})=\frac{\sqrt{\pi}}{k_{pp}}f_{pp}^\textrm C({\bf q}) S_d({\bf q}/2),
\end{eqnarray}
where $f_{pp}^\textrm C$ is the antisymmetric Coulomb  amplitude of the $pp$ scattering
\cite{MS}:
\begin{eqnarray}
\label{ppantisym}
f_{pp}^\textrm C({\bf q})=f(\theta_{pp})-\frac{1}{2}(1+\bfg\sigma \cdot \bfg \sigma_p)
f(\pi-\theta_{pp})
\end{eqnarray}
with
\begin{eqnarray}
\label{kulonpp}
f(\theta_{pp})= -\frac{\alpha}{4vk_{ p p}\sin^2{\theta_{pp}/2}}\exp{\left\{ -i\frac{\alpha}{v}
\ln{\sin{\frac{\theta_{pp}}{2}}+2i\chi_0} \right \}},
\end{eqnarray}
here $\alpha$  is  the fine structure constant,
 $v$ ($k_{pp}$)  is the velocity (momentum)  of the proton
  in the cms  $pp$ system, $\chi_0$ is the Coulomb phase. The momentum ${\bf q}$ transferred in the
 process $pd\to pd$  is related to the $pp$ scattering angle
  $\theta_{pp}$ in the  $pp$ cms  as  $q=2k_{pp}\sin\theta_{pp}/2$.

 In Eq. (\ref{kulonpd}) $S_d({\bf q}/2)$  is the elastic form factor of the deuteron which
can be presented in the form \cite{PK}
\begin{eqnarray}
\label{dff}
S_d({\bf q}/2)= S_0(q/2) -\frac{1}{\sqrt{8}}S_2(q/2)S_{12}({\hat {\bf q}};{\bfg\sigma}_p,{\bfg\sigma}_n).
\end{eqnarray}
Here
\begin{eqnarray}
\label{T2}
 S_{12}({\hat{\bf q}};{\bfg\sigma}_p,{\bfg\sigma}_n)
=3({\bfg\sigma}_p\cdot {\hat{\bf q}})({\bfg\sigma}_n\cdot {\hat{\bf q}})-
{\bfg\sigma}_p\cdot{\bfg\sigma}_n
 \end{eqnarray}
 is the tensor operator,
  ${\bfg\sigma}_n ({\bfg\sigma}_p)$ are the Pauli matrices acting on the spin states of the neutron
 and proton in the deuteron, ${\hat{\bf q}}$  is the unit vector
  directed along the transferred momentum ${\bf q}$.
  The form factors $S_0$ and  $S_2$  are related to the   $S$-  and  $D$- components of the
 deuteron wave function \cite{PK}:
\begin{eqnarray}
S_0(q)=S_0^{(0)}(q)+S_0^{(2)}(q), S_2(q)=S_2^{(1)}(q)+S_2^{(2)}(q),
\label{s0s2}
\end{eqnarray}
here
\begin{eqnarray}
S_0^{(0)}(q)=\int_0^\infty dr u^2(r)j_0(qr),\,\, S_0^{(2)}(q)=\int_0^\infty dr w^2(r)j_0(qr),
\nonumber \\
S_2^{(1)}(q)=2\int_0^\infty dr u(r)w(r)j_2(qr),\,\,
S_2^{(2)}(q)=-\frac{1}{\sqrt{2}}\int_0^\infty dr w^2(r)j_2(qr).
\label{ssds}
\end{eqnarray}

% Since the Coulomb amplitude
%(\ref{kulonpd})  does not depend on the  spins of the protons  it is diagonal in respect
% of spins of protons ($\mu$) and the deuteron ($\lambda$).

\subsection{TVPC $pN$ scattering amplitudes}
In general case  TVPC NN interaction  contains 18 different terms~\cite{herczeg}.
 We  consider here only the
following terms of the t-matrix  of the  elastic $pN$ scattering investigated
%which were under discussion
 in  Ref.~\cite{beyer}:
\begin{eqnarray}
\label{TVNN}
t_{pN}= f_N({\bfg \sigma\cdot \bfg\sigma}_N)({\bf q}\cdot {\bf k})/m_p^2+
{h_N[({\bfg \sigma} \cdot {\bf k})({\bfg \sigma}_N \cdot {\bf q})+
({\bfg \sigma}_N \cdot {\bf k})({\bfg \sigma} \cdot {\bf q})-
\frac{2}{3}({\bfg \sigma}_N \cdot{\bfg \sigma})
({\bf k}\cdot {\bf q}) ]}/m_p^2
+ \\ \nonumber
+g_N [{\bfg \sigma} \times {\bfg \sigma}_N]\cdot [{\bf q }\times{\bf k}]/m_p^2
+{g_N^\prime ({\bfg \sigma} - {\bfg \sigma}_N)\cdot i\,[{\bf q}\times {\bf k}]
[{\bfg \tau} \times{\bfg \tau}_N]_z}/m_p^2.
\end{eqnarray}
Here ${\bfg \sigma}$
(${\bfg \sigma}_N$)
 is the Pauli matrix acting on the spin state of the proton
(nucleon $N=p,n$),
${\bfg \tau}$
(${\bfg \tau}_N$)
 is the isospin matrix acting on the  isospin state of the proton (nucleon),
 $\bf q=\bf p- \bf  p'$.
 In the framework  of the   phenomenological meson exchange interaction
 the term $g_N^\prime$ corresponds to the $\rho-$meson exchange,
 and  the $h_N$-term is caused by the axial meson exchange. As shown in Ref. \cite{simonius75},
 the contribution of the $\pi$- and $\sigma$-meson exchanges to TVPC NN interactions is excluded.
% In the framework  of the   phenomenological meson exchange interaction
% the term $g'$ corresponds to $\rho$-meson exchange,
%and $h$-term provides the axial meson $h_1$ exchange.
% with
%$h_N=-i2G_{h}{\bar G}_{h} F(q^2)(m_h^2+{\bf q}^2)^{-1}$,
%where $G_{h}$ (${\bar G}_{h}$) is the ordinary (TVPC) $hNN$ coupling constant,
%$m_h$ is the h-meson mass and $F(q^2)$ is the $hNN$ vertex
%form factor \cite{beyer}.
The TVPC NN interaction potential corresponding to $h_1(1170)$, $I^G(J^{PC})=0^-(1^{+-})$
exchange in r-space
has a form \cite{lazauskas}
\begin{eqnarray}
\label{VNNh}
V_h({\bf r})=-\frac{G_h{\bar G}_h m_h^2}{2\pi m_N^2} {Y_1(x)}\left [({\bfg \sigma}_1{\bar {\bf p}})
({\bfg \sigma}_2{\hat {\bf r}}) +({\bfg \sigma}_2{\bar {\bf p}})
({\bfg \sigma}_1{\hat {\bf r}})\right ],
\end{eqnarray}
 where  $G_{h}$ (${\bar G}_{h}$) is the ordinary (TVPC) $hNN$ coupling constant,
$m_h$ is the $h_1$-meson mass, ${\bar {\bf p}}=({\bf p}_1-{\bf p}_2)/2$,\, ${\bf r}={\bf r}_1-{\bf r}_2$,
 $Y_1(x)= (1+x)e^{-x}/x^2$,  $x= m_h r$; ${\bfg \sigma}_i/2$ and
 ${\bf p}_i$ are the spin-operator and momentum of {\it i}-th nucleon, respectively,
 and ${\bf r}_i$ is its r-coordinate $(i=1,2$). Making the Fourier-transformation of Eq. (\ref{VNNh})
 we obtain
 the interaction potential in p-space and, therefore, find the factor $h_N$ in Eq. (\ref{TVNN})
% one can find for the $h$-term
 within the Born approximation for
the $t_{pN}$ amplitude as the following
%\begin{eqnarray}
%\label{VNNh1}
%<{\bf p}'|V_h({\bf r})|{\bf p}>=\frac{1}{2}<{\bf p}'|V_h({\bf r})+V_h^+({\bf r})|{\bf p}>=
%-i\frac{2G_h{\bar G}_h }{ m_N^2} \frac{F_{hNN}(q^2)}{m_h^2+{\bf q}^2}
%\left [({\bfg \sigma}_1{ {\bf k}})
%({\bfg \sigma}_2{\bf q}) +({\bfg \sigma}_2{ {\bf k}})
%({\bfg \sigma}_1{\bf q})\right ],
%\end{eqnarray}
%where ${\bf k}={\bf p}+{\bf p}'$, ${\bf q}={\bf p}-{\bf p}'$,
%$F_{hNN}(q^2)=(\Lambda^2-m_h^2)/(\Lambda^2+{\bf q}^2)$
%is the monopole  $hNN$  form factor.
%Here we used that $V_h({\bf r})$ operator is Hermite:$V_h=V_h^+$.
%  From comparison of Eq. (\ref{VNNh1}) with
%Eq. (\ref{TVNN}), one can find for the $h$-term  within the Born approximation for
%the $t_{pN}$ amplitude
\begin{eqnarray}
\label{hform}
h_N
=-i\phi_h\frac{2G_h^2}{m_h^2+{\bf q}^2}F_{hNN}({\bf q}^2),
\end{eqnarray}
where $\phi_h={\bar G}_{h}/G_{h}$ is the strength of the T-invariance violating potential
of $h_1$-meson exchange relative to the T-conserving one, and
$F_{hNN}(q^2)={(\Lambda^2-m_h^2)}/{(\Lambda^2+{\bf q}^2)}$
is the phenomenological monopole formfactor in the $hNN$ vertex.
% We should note that $h_N$ in
% Eqs.(\ref{VNNh1}),
%Eq. (\ref{hform}) differs from that in Ref.\cite{beyer} by the numerical
% factor of 32.
 Similarly proceeding from the   TVPC $\rho$-meson exchange NN-potential
in r-space \cite{lazauskas,haxton93,Engel94} we  find for the $g'$-term in Eq. (\ref{TVNN})
% however, we skip them in view of vanishing contribution of the $\rho$- meson to $\widetilde g$.
\begin{eqnarray}
\label{gsform}
g^\prime_N=-\phi_\rho\frac{1}{2}\frac{g^2_\rho\kappa}{m_\rho^2+{\bf q}^2}F_{\rho NN}({\bf q}^2),
 \end{eqnarray}
where $\phi_\rho={\bar g}_\rho/g_\rho$ is the ratio of the TVPC $\rho NN$ coupling constant ${\bar g}_\rho$
 to the strong one  $g_\rho$, $m_\rho$ is the mass of the $\rho$-meson, $\kappa$ is the anomalous
magnetic moment of the nucleon, $F_{\rho NN}$ is the $\rho NN$ vertex formfactor.

\subsection{The Glauber operator}

The Glauber operator
of the elastic $pd$ scattering in general case
can be written  as
\begin{eqnarray}
\label{18total1}
%\begin{array}{l}
M({\bf q}, {\bf Q}; {\bf S}, {\bfg \sigma})=
\iiint{e}^{i{\bf Q}{\bf r}}{\Psi }_{d}^+({\bf r}){\mathrm O}{\Psi }_d({\bf r}) {d^3} r,
\end{eqnarray}
where $ {\Psi }_d$ is the deuteron wave function,
  ${\bf q}$ is the transferred momentum, $\bf S=({\bfg \sigma}_p+{\bfg \sigma}_n)/2$
is the spin-operator of the deuteron nucleons, $\bfg \sigma/2$ is the spin-operator of the incoming
 proton.
%Here for  the deuteron we
 We use the deuteron
 wave function generated by the NN interaction which conserves  time reversal invariance
 and P-parity and has the following  standard form
% corresponding to conservation of the time reversal invariance and P-parity
 \begin{equation}
\label{WF}
\Psi _d \left( {\bf r}; \bfg \sigma_n, \bfg \sigma_p \right)
=\frac{1}{\sqrt{4\pi }r}\left( u\left( r \right)+\frac{1}{\sqrt{8}}w
\left( r \right)\cdot {{S}_{12}}\left( {\bf{\hat{r}}};\bfg \sigma_n,\bfg\sigma_p \right) \right),
\end{equation}
where lower index    $n$  ($p$)
 refers to the neutron  (proton) of the deuteron target;
$u$  and  $w$ denote the
 S- and  D- wave of the deuteron, respectively; the tensor operator
${{S}_{12}}\left( {\bf{\hat{r}}};\bfg \sigma_n,\bfg\sigma_p \right)$
 is defined by Eq. (\ref{T2}).
%\begin{eqnarray}
%{{S}_{12}}\left( \hat{\bf{r}};{{{\bfg{\sigma }}}_{n}},{{{\bfg{\sigma }}}_{p}} \right) =
%3({\bfg\sigma}_p\cdot {\hat{\bf r}})({\bfg\sigma}_n\cdot {\hat{\bf r}})-{\bfg\sigma}_p\cdot{\bfg\sigma}_n.
%\label{tensoroper}
%\end{eqnarray}

The  operator for the single and double scattering mechanisms of $pd$ scattering
in general case
(beyond the collinear kinematics)
can be written  as an expansion over the
Pauli matrices $\bfg \sigma_n,\, \bfg \sigma_p$ and  in notations of Ref. \cite{PK} takes the form
\begin{equation}
\label{operator}
{\mathrm O}\left( \bfg{\sigma },{{{\bfg{\sigma }}}_{n}},{{{\bfg{\sigma }}}_{p}} \right)=
U\left( {\bfg{\sigma }} \right)+{{\bf{V}}_{n}}\left( {\bfg{\sigma }} \right)\cdot {{\bfg{\sigma }}_{n}}+
{{\bf{V}}_{p}}\left( {\bfg{\sigma }} \right)\cdot {{\bfg{\sigma }}_{p}}+
{{W}_{ij}}\left( {\bfg{\sigma }} \right)\cdot \left( {{\sigma }_{ni}}{{\sigma }_{pj}}+
{{\sigma }_{nj}}{{\sigma }_{pi}} \right),
\end{equation}
 here   $i, j =  q,\ n, k$ are indices  of the projections onto
directions of three orthogonal vectors  ${\hat{\bf q}},{\hat {\bf n}},
{\hat {\bf k}}$ introduced in  Eq. (\ref{pnamp}).
 The operators  $U$, $V$, $W$ act only on the spin state of the beam proton
  and do not depend on the spins and  coordinates ${\bf r}$ of the target
  nucleons.
%When sandwiching  the operator (\ref{operator}) between
%the deuteron wave functions (\ref{WF})
When making  the matrix element of the operator (\ref{operator}) over
the deuteron wave functions (\ref{WF})
 we obtain from Eq. (\ref{18total1})
\begin{eqnarray}
\label{18total}
M({\bf q}, {\bf Q}; {\bf S}, {\bfg \sigma})=
\iiint{e}^{i{\bf Q}{\bf r}}{\Psi }_{d}^+({\bf r}){\mathrm O}{\Psi }_d({\bf r}) {d^3} r  \nonumber \\
   =U{S}_{0}+{\bf V} {\bf S}S_{0}^{\left( 0 \right)}
+\left [ {W_{ij}}\left\{ {S}_i,{S_j} \right\}-W_{ii} \right ] S_{0}^{\left( 0 \right)}
-\frac{1}{\sqrt{2}}U{S}_{12}( {\hat{\bf Q}};{\bf S},{\bf S}  )S_2
 \nonumber \\
-\frac{1}{\sqrt{8}}{S}_{12} ( {\hat{\bf Q}};{\bf V},{\bf S}  )
S_{2}^{ ( 1 )}
 +\sqrt{8}\, {{W}_{ii}} {{S}_{12}} ( \hat{\bf{Q}};{\bf S},{\bf S} )S_{2}^{( 1 )}
  -\frac{1}{\sqrt{2}}{{S}_{12}} ( \hat{\bf {Q}};{\bf V},{\bf S} )S_{2}^{\left( 2 \right)}
  -\frac{1}{2}{\bf V}{\bf S}S_{0}^{\left( 2 \right)}+  \nonumber \\
 -{{W}_{ii}} {{S}_{12}}( {\hat{\bf Q}};{\bf S},{\bf S}  )
 S_{2}^{\left( 2 \right)}-2\,{{W}_{ii}}S_{0}^{\left( 2 \right)}- \nonumber \\
  -\sqrt{2}\, W_{ij}\,\left[ \left \{ {S}_i,S_j \right \}S_{12}(
  \hat{\bf Q};{\bf S},{\bf S}  )
  +{{S}_{12}}( \hat{\bf Q};{\bf S},{\bf S} )\left\{ {{S}_{i}},{{S}_{j}} \right\} \right ]
S_{2}^{\left( 1 \right)}+ \nonumber \\
+\frac{1}{16\pi }{{W}_{ij}}\int{{{d}^{3}}r
\frac{1}{r^2}{{e}^{i{\bf Q}{\bf r}}}{{w }^{2}}}\,
  {{S}_{12}}\left( \hat{\bf r};{{{\bfg \sigma }}}_{n},
{{{\bfg{\sigma }}}_{p}} \right)\left\{ {{S}_{i}},{{S}_{j}}
 \right\}{{S}_{12}}\left( {\bf \hat{r}};{{{\bfg \sigma }}_{n}},{{{\bfg \sigma }}_{p}} \right).
\end{eqnarray}
Here  we use the notations
%${\bf S}=({\bfg \sigma}_n+{\bfg \sigma}_p)/2$,
 ${\bf V}={\bf V}_p+{\bf V}_n$ and  $\{S_i,S_j\}=S_iS_j+S_jS_i$; the form factors
$S_0(Q), S_2(Q), S_0^{(0)}(Q),S_0^{(2)}(Q), S_2^{(1)}(Q)$ and $S_2^{(2)}(Q)$ are defined in
 Eqs. (\ref{s0s2}), (\ref{ssds}); the tensor operators
 $S_{12}({\hat{ \bf Q}}; {\bf V}, {\bf S})$ and  $S_{12}({\hat {\bf Q}}; {\bf S}, {\bf S})$ are defined
 in Eq. (\ref{T2}).
 In Eq. (\ref{18total}) summation is performed over repeating indexes  $i,j$.
To make the integration over directions of the vector  $\bf r$ in  Eq. (\ref{18total}),
 we used the following relation \cite{fealdt}
\begin{eqnarray}
\label{faeld}
\iint d\Omega_{\bf r}\exp{(-i\bf Q\bf r)}T_l(\hat {\bf r})= 4\pi j_l(Qr) (-i)^l T_l(\hat {\bf Q}),
\end{eqnarray}
where $j_l(x)$ is the spherical Bessel function,
 $T_2(\hat {\bf n})=(\bfg \sigma_p\cdot \hat {\bf n})(\bfg \sigma_n\cdot \hat {\bf n})-
\frac{1}{3}(\bfg \sigma_p\cdot\bfg \sigma_n)$, $T_0(\hat {\bf n})=\bfg \sigma_p\cdot \bfg \sigma_n$;
$\hat {\bf n}$, $\hat {\bf Q}$ and  $\hat {\bf r}$ are unit vectors along ${\bf n}$, ${\bf Q}$ and
$ {\bf r}$, respectively.

 Eq. (\ref{18total}) is a generalization of Eq. (18) from Ref. \cite{PK}.
 The difference from Ref. \cite{PK} consists in two  following respects. First,
 the operators $U$, ${\bf V}$, $W_{ij}$ contain not only T-even P-even terms, but T-odd P-even terms
 as well.
 Second,  we present in
 Eq. (\ref{18total}) all terms allowed within the Glauber theory, whereas in Ref.\cite{PK}
 small spin-dependent terms (of the order higher than two in definitions of Ref. \cite{PK})
 were neglected. These terms are small at high energies about 1 GeV, however, may be
important at lower energies $\sim 100$ MeV,
 corresponding the COSY experiment \cite{TRIC}.

\subsection{ Differential spin observables}
 Vector analyzing powers $A_y$ and  spin correlation coefficients $C_{i,j}$, $C_{ij,k}$
of the elastic $pd$ scattering are calculated as 
\begin{eqnarray}
\label{cijk}
A^p_y=TrM\sigma_yM^+/Tr MM^+, \, A^d_y=TrM S_yM^+/Tr MM^+, \nonumber \\
C_{xz,y}=TrMS_{xz}\sigma_yM^+/Tr MM^+, \, C_{y,y}=TrMS_y\sigma_yM^+Tr MM^+, \nonumber \\
C_{x,z}=TrM S_x\sigma_z M^+/Tr MM^+, \,C_{z,x}=TrM S_z\sigma_x M^+/Tr MM^+,
\end{eqnarray}
where $M$ is the transition operator given by Eq. (\ref{18total1}).
 Details of these calculations in terms of invariant amplitudes are described in Ref. \cite{TUZyaf}. 

\section{Null-test signal of TVPC interactions}

Eq. (\ref{18total}) gives the single scattering $pd$  amplitude
%  $M^{(s)}$
 if one put
 ${\bf Q}={\bf q}/2$, where $\bf q$ is the momentum transferred in the $pd$ scattering.
Within the Glauber theory the amplitude of the single scattering mechanism is proportional
 to the on-shell $t_{pN}({\bf q})$ amplitude.
 At zero
 scattering angle  the TVPC amplitude (\ref{TVNN}) vanishes, therefore,
 the corresponding  $pd$-scattering amplitude and total
 cross section $\widetilde \sigma$ of $pd$-scattering are equal to zero in the single
 scattering Glauber
  approximation. Furthermore,
 the  $f$-term in Eq. (\ref{TVNN}) gives zero contribution within the Glauber theory both for the single and
 double scattering, because
 for the on-shell $pN$ scattering involved into multistep scattering
% Glauber operator
 (\ref{operator}),
 one has $({\bf q}\cdot {\bf k})=0$. For the same reason, the component of the $h$-term proportional to
${\bfg \sigma}_N {\bfg \sigma}$  vanishes in Eq. (\ref{TVNN}) too.
 The rest $g^\prime$, $g$ and $h$ terms  contribute to
 the double scattering forward $pd$-elastic amplitude.

The double scattering amplitude  is given   by integration of
%the spin-matrix element of the operator given by
Eq.(\ref{18total}) over $\bf Q\equiv{\bf q}'$ \cite{PK}
\begin{eqnarray}
\label{DS}
M^{(d)}=\frac{i}{2\pi^{3/2}}\iint d^2q'M({\bf q},{\bf q}'; {\bf S}, {\bfg \sigma}).
\end{eqnarray}
According to Eq. (\ref{me}), in order to get
the TVPC $\widetilde g$ amplitude one has  to calculate  the matrix element of
the operator given by  Eq.(\ref{DS})
at $\bf q=0$ over
 definite initial $|\mu,\lambda>$ and final $|\mu',\lambda'>$
 spin states:
%  obtained  by integration of the spin-matrix element of the operator given by
%Eq.(\ref{18total}) over $\bf Q\equiv{\bf q}'$.
%Below we calculate   the   $pd$ elastic scattering  amplitude
% at zero scattering angle for the double scattering mechanism
%normalized according to Ref. \cite{PK}
\begin{eqnarray}
{\widetilde g}=\frac{1}{(2\pi)^{3/2}}\int d^2 q^\prime
 <\mu'=\frac{1}{2},\lambda'=0|M({\bf q}=0,{\bf q}^\prime; {\bf S}, {\bfg \sigma})|
\mu=-\frac{1}{2},\lambda=1>.
% M^{(d)}= \frac{i}{2\pi^{3/2}}\int d^2 q'M({\bf q}').
\label{g5ds}
\end{eqnarray}
When considering the double scattering  mechanism,
in addition to three vectors $\{\hat {\bf k}, \hat {\bf q},\hat { \bf n} \}$ defined
after Eq. (\ref{pnamp})
 it is convenient
 to introduce two more sets of orthonormal unit vectors
$\{\hat {\bf k}_j, \hat {\bf q}_j,\hat { \bf n}_j \}$ for the first ($j=1$) and second
$(j=2)$ collision as it was done in Ref. \cite{PK}.
 At zero scattering angle we have ${\bf q}_1=-{\bf q}_2=-{\bf q}'$, where ${\bf q}_1$
 (${\bf q}_2$) is the transferred momentum in the first (second) collision;
${\bf n}_1=-[{\bf k}\times {\bf q}']$, ${\bf n}_2=[{\bf k}\times {\bf q}']$,
${\bf k}_1={\bf k}_2={\bf k}+{\bf q}'$. In the eikonal approximation vectors $\bf q$
and ${\bf q}'$ are orthogonal to $\bf k$ and we assume  ${\bf k}_1={\bf k}_2={\bf k}$
\cite{PK}. The Cartesian projections for unit vectors  can be written in terms of components
of the two-dimensional  vector ${\bf q}'=(q'_x, q'_y)$ (OZ axis is directed along ${\bf k}$)
 as ${\hat n}_{1x}=q'_y/q'$,
${\hat n}_{1y}=-q'_x/q'$, ${\hat n}_{1z}=0$.

\begin{figure}[t]
 \includegraphics[width=0.50\textwidth,angle=0]{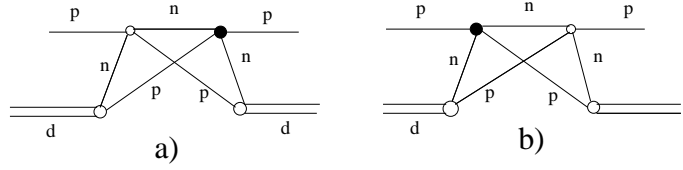}
\caption{ Double scattering mechanism with TVPC (black circle) and T-even P-even (open circle)
charge-exchange $pn$ interaction.
}
\label{fig1}
\end{figure}

\subsection{$g^\prime$ term}

Non-zero matrix elements of the isospin-operator connected with
 the $g'$ term in Eq. (\ref{TVNN}) are
\begin{eqnarray}
\label{isospin}
<n,p|[{\bfg \tau} \times{\bfg \tau}_N]_z|p,n>=-i2,\,\,
<p,n|[{\bfg \tau} \times{\bfg \tau}_N]_z|n,p>=i2.
\end{eqnarray}
Therefore, the $g'$ term contributes only to the  charge exchange transitions.
 Two allowed double scattering amplitudes   with one TVPC and another one T-even P-even
 $pN$-interaction
are depicted in Fig. \ref{fig1}. Within the operator formalism  these two terms can be
evaluated in the following way.
For pure  T-even P-even (TCPC) interactions the transition operator
 for the charge-exchange mechanism of
 the $pd\to pd$  process
 has a form \cite{glauber-franco}
 \begin{eqnarray}
\label{ce-oper}
%O^c=\frac{1}{2}[M_p({\bf q}_2)t^c_{pn}({\bf q}_1)+t^c_{np}({\bf q}_2)M_n({\bf q}_1)].
O^c_{TVPC}=-\frac{1}{2}[M_c({\bf q}_2)M_c({\bf q}_1)],
\end{eqnarray}
 where $M_c({\bf q})=M_n({\bf q})-M_p({\bf q})$ is the transition  operator for the strong
 charge-exchange
 $pn\to np$  amplitude that is  equal to the $np\to pn$ amplitude. We note that according
 to Eq. (\ref{isospin}), for the
TVPC NN-interaction with the $g^\prime$ terms the amplitude $pn\to np$ differs from
the amplitude  $np\to pn$ by the sign. In order to get the TVPC operator of the  charge-exchange
 $pd$ scattering, we
 make in Eq. (\ref{ce-oper}) the replacement
%$M_c({\bf q})\to M_c({\bf q})+t_c({\bf q})$,
$M_c({\bf q})\to M_c({\bf q})+T_c({\bf q})$,
where  the index $c$ means either $pn\to np$ or $np\to pn$ and
  $T_c$ is the TVPC charge-exchange NN-scattering operator, normalized as  $M_N$ in Eq. (\ref{pnamp}) and
 related to the $t_{pN}$-
operator  given by Eq. (\ref{TVNN}) as
\begin{eqnarray}
\label{TPK}
T_{pN}=\frac{m_N}{4\sqrt{\pi}k_{pN}}t_{pN}.
\end{eqnarray}
Furthermore, we neglect the terms of the second order in $T_c$ as compared to the first order
 and omit the T-even  term $\sim M_cM_c$. As a result, the TVPC  charge-exchange operator
takes the form
\begin{eqnarray}
\label{ce-oper-tvpc}
O^c_{TVPC}=-\frac{1}{2}[M_{np\to pn}({\bf q}_2)T_{pn\to np}({\bf q}_1)+T_{np\to pn}({\bf q}_2)
M_{pn\to np}({\bf q}_1)].
\end{eqnarray}
For further evaluation it is convenient to use $M_{np\to pn}=M_{pn\to np}=M_n-M_p$.
Under the sign of the integral over ${\bf q}'$ in Eq. (\ref{g5ds})
the operator  (\ref{ce-oper-tvpc}) is not  changed after the
substitution ${\bf q}_1\leftrightarrow {\bf q}_2$. In order to find the  operators $U$,
${\bf V}_p$, ${\bf V}_n$, and $W_{ij}$ introduced in  Eq. (\ref{operator}),
 it is convenient to add
%for symmetry
to the right side of
 Eq. (\ref{ce-oper-tvpc})
 the term  $O^{c}_{TVPC} (1\leftrightarrow2)$ and divide the obtained sum by
 the factor of 2:
\begin{eqnarray}
\label{gsoper}
O^c_{TVPC}= f_I+f_{II},
\end{eqnarray}
where
\begin{eqnarray}
\label{fI}
f_I=-\frac{1}{4}[(M_n({\bf q}_2)T_{pn\to np}({\bf q}_1)+T_{np\to pn}({\bf q}_2)
M_n({\bf q}_1)) +({\bf q}_1 \leftrightarrow{\bf q}_2) ], \nonumber \\
f_{II}=\frac{1}{4}[M_p({\bf q}_2)T_{pn\to np}({\bf q}_1)+T_{np\to pn}({\bf q}_2)
M_p({\bf q}_1)
+({\bf q}_1 \leftrightarrow{\bf q}_2) ].
\end{eqnarray}

Using Eqs. (\ref{pnamp}), (\ref{isospin})
 and symmetry in respect of the replacement
$({\bf q}_1 \leftrightarrow{\bf q}_2)$, we find that $A_N$, $B_N$, $G_N$,
and $H_N$  terms cancel in operators $f_I$ and $f_{II}$:
\begin{eqnarray}
\label{fI2}
f_I=\frac{g'}{\Pi}\biggl [C_n (\bfg\sigma\cdot {\hat{\bf n}}_1)
({\bfg\sigma}_n-{\bfg\sigma}_p)\cdot {\bf n}_1 -
C^\prime_n ({\bfg \sigma}_n\cdot {\hat {\bf n}}_1)
({\bfg\sigma}_p)\cdot {\bf n}_1) +C^\prime _n{\bf n}_1{\hat {\bf n}}_1\biggr ],\nonumber \\
\label{fII2}
f_{II}=\frac{g'}{\Pi}\biggl [C_p ({\bfg\sigma}\cdot {\hat {\bf n}}_1)
({\bfg\sigma}_p-{\bfg\sigma}_n)\cdot {\bf n}_1 -
C^\prime_p ({\bfg \sigma}_p\cdot {\hat {\bf n}}_1)
({\bfg\sigma}_n)\cdot {\bf n}_1) +C^\prime _p{\bf n}_1 {\hat {\bf n}}_1\biggr ],
\end{eqnarray}
where
\begin{eqnarray}
\Pi={4\sqrt{\pi}m_Nk_{pN}}.
\end{eqnarray}
%
%Using Eqs.(\ref{isospin}) and
Making the sum $f_I+f_{II}$
we find  the operators $U,\,V,\,$ and $W_{ij}$ in Eq. (\ref{operator})
 for the $g^\prime$-term as the following
\begin{eqnarray}
\label{gprime-uvw}
U=\frac{g'}{\Pi} (C_n^\prime+C_p^\prime){\bf n}_1 {\hat {\bf n}}_1, \nonumber \\
%\nonumber
{\bf V}_p=(C_p-C_n)(\bfg \sigma\cdot {\bf n}_1){\hat {\bf n}}_1,\,
{\bf V}_n=(C_n-C_p)(\bfg \sigma\cdot {\bf n}_1){\hat {\bf n}}_1, \nonumber \\
% \nonumber
W_{ij}=-\frac{g'}{\Pi}(C_n^\prime+C_p^\prime) {n_1}_i {\hat n}_{1 {\small j}},\,
W_{ii}=-\frac{g'}{\Pi}(C_n^\prime+C_p^\prime) {n_1} {\hat {n}_1}.
 \end{eqnarray}

 One can see from Eqs.~(\ref{gprime-uvw}), that ${\bf V}={\bf V}_p+{\bf V}_n\equiv 0$.
Furthermore, taking into account the relation
${\bf V}_p{\bfg\sigma}_p+{\bf V}_n{\bfg\sigma}_n=
{\bf V}{\bf S}$,
 we find that  for the $g^\prime$-term the operator Eq.~(\ref{operator})
 does not depend on
 the spin of the
 proton beam $\bfg \sigma$. As a result the transition operator  given by  Eq.~(\ref{18total})
 is diagonal in respect of spins of the proton beam.
  According to Eq. (\ref{me}), it means that the contribution of $g^\prime$ term
 to the TVPC amplitude $\widetilde g$ is equal to zero.
 We emphasize that this result is true   for the S- and D- components of the
 deuteron wave function and for all spin terms in the transition amplitude (\ref{18total})
 allowed in the Glauber formalism.
 % including those which
 %were neglected in Ref.\cite{PK}.
  It is easy to find that
 this result is valid for the $nd$  scattering too.

\subsection{$h$ and $g$-terms}
\label{ghterms}

%Four double scattering terms are allowed in the first order over
% since
The TVPC interaction  corresponding to the $h$ and  $g$
 terms in Eq. (\ref{TVNN}) occurs both in the $pp$- and $pn$- elastic scattering.
Following to
Ref. \cite{glauber-franco} (see Eq. (2.7) in it) and \cite{PK}
% we
%split the transition operator for the double scattering $Q^{(d)}$ into symmetric
% $O^{(d)}_{+}$ and antisymmetric
% $O^{(d)}_{-}$ parts
we consider the symmetric
 $O^{(d)}_{+}$ and antisymmetric
 $O^{(d)}_{-}$ parts
of the operator $Q^{(d)}$:
\begin{eqnarray}
\label{oplus}
Q^{(d)}_{+}=\frac{1}{2}[(T_{pp}({\bf q}_1)+M_p({\bf q_1})(T_{pn}({\bf q}_2)+M_n({\bf q}_2))
 +(T_{pn}({\bf q}_2)+M_n({\bf q}_2)) (T_{pp}({\bf q}_1)+M_p({\bf q_1})],\nonumber \\
\label{opminus}
Q^{(d)}_{-}=\frac{1}{2}[(T_{pp}({\bf q}_1)+M_p({\bf q_1})(T_{pn}({\bf q}_2)+M_n({\bf q}_2))
 -(T_{pn}({\bf q}_2)+M_n({\bf q}_2)) (T_{pp}({\bf q}_1)+M_p({\bf q_1})].
\end{eqnarray}
According to Refs.\cite{glauber-franco} and \cite{PK} the matrix elements of the antisymmetric
 operator
 $Q^{(d)}_{-}$ over the spin $S=1$ states of
 the deuteron are negligible and therefore, we drop this term.
  In the operator $Q^{(d)}_{+}$  one can neglect  terms of
 the second order
 in the TVPC interaction ($\sim T_{pp}T_{pn}$) as compared to the first order $T_{pN}$ 
 and should  omit the pure T-even
 terms $\sim M_{p}M_{n}$. Thus, the double scattering TVPC operator consists of the following
four terms
\begin{eqnarray}
\label{doplus}
Q^{(d)}_{+}=\frac{1}{2}[T_{pp}({\bf q}_1)M_n({\bf q}_2)+M_n({\bf q}_2)T_{pp}({\bf q}_1)+
T_{pn}({\bf q}_2)M_p({\bf q}_1)+M_p({\bf q}_1)T_{pn}({\bf q}_2)].
\end{eqnarray}
%Under the sign of the integral over ${\bf q}'$ in Eq. (\ref{g5ds})
%this operator does not change after the
%substitution ${\bf q}_1\leftrightarrow {\bf q}_2$.
As in the case of $g^\prime$-term,
%In order to find operators $U$,
%${\bf V}_p$, ${\bf V}_n$, $W_{ij}$ introduced in  Eq. (\ref{operator}),
 it is convenient to add
%for symmetry
 to the right side of Eq. (\ref{doplus})
 the term  $O_{+}^{(d)}(1\leftrightarrow2)$ and divide the obtained sum by
 the factor of 2.

{\it For the $h$-term}
we find  the operators $U,\,V$, and $W_{ij}$ in Eq.(\ref{operator}) for the forward
  double scattering as 
\begin{eqnarray}
U=0,\, {\bf V}_p=0,\, {\bf V}_n=0,\, W_{ii}=0, \\ \nonumber
W_{ij}=\frac{1}{\Pi} \Biggl \{C^\prime_n(q')h_p(q') [{\bfg \sigma}{\bf k}_1
{\hat n}_{2i}q_{1j}+
{\bfg \sigma}{\bf q}_1 k_{1j}{\hat n}_{2i}]+
C^\prime_p(q')h_n(q')[{\bfg \sigma}{\bf k}_1{\hat n}_{2j}q_{1i}+
{\bfg \sigma}{\bf q}_1 k_{1i}{\hat n}_{2j}]\Biggr \}, \\
 W_{ij}\{S_i,S_j\}=\frac{C^\prime_nh_p+C^\prime_ph_n}{\Pi}\times \nonumber \\
\left
\{({\bfg\sigma}{\bf k}_1)[({\hat {\bf n}}_2{\bf S})({\bf q}_1{\bf S})+
({\bf q}_1{\bf S})({\hat {\bf n}}_2{\bf S})]+
({\bfg\sigma}{\bf q}_1)[({\bf k}_1{\bf S})({\hat {\bf n}}_2{\bf S})+
({\hat {\bf n}}_2{\bf S})({\bf k}_1{\bf S})]
\right
%\biggr
\}.
\label{h}
\end{eqnarray}
In Eq. (\ref{operator})
in this case  only the operator $W_{ij}$ depends on
 the  beam proton  spin and, therefore, gives non-zero contribution to ${\widetilde g}$.
The matrix elements over the proton spin states are
\begin{eqnarray}
<\mu'=+\frac{1}{2}|W_{ij}\{S_i,S_j\}|\mu=-\frac{1}{2}>=\nonumber \\
(C^\prime_nh_p+C^\prime_ph_n)(-q'_x+iq'_y)\Biggl [S_zS_xq'_y-S_zS_yq'_x
-q'_yS_xS_z+q'_xS_yS_z\Biggr ]\frac{k^2}{\Pi|{\bf n}_1|}.
\label{hme}
\end{eqnarray}
For the deuteron spin matrix elements of the operator $M$
one has $<\lambda'=0|M|\lambda=1>=-(M_{zx}+iM_{zy})/\sqrt{2}$.
Finally the spin  matrix element in the S-wave approximation is the following
\begin{eqnarray}
<\mu'=\frac{1}{2},\lambda'=0|M({\bf q}=0,{\bf q}^\prime; {\bf S}, {\bfg \sigma})|
\mu=-\frac{1}{2},\lambda=1>=
-\frac{i k}{\sqrt{2}} \frac{(C^\prime_nh_p+C^\prime_ph_n)q'}{\Pi}S_0^{(0)}(q').
\label{hmed}
\end{eqnarray}
%
%The D-component of the deuteron wave function   could contribute due to its interference with
% the S-wave  in the terms containing the $S_2^{(1)}$ form factor in Eq. (\ref{18total}).
% However, the
% matrix elements of the corresponding operators $S_{12}({\bf Q}; {\bf S},{\bf S})$ and
% $S_{12}({\bf Q}; {\bf V},{\bf S})$ over the deuteron
% spin states are zeros:
% $<\lambda'=0|S_{12}({\bf Q}; {\bf S},{\bf S}) |\lambda=+1>\equiv0$,\,
%$S_{12}({\bf Q}; {\bf V}=0,{\bf S})\equiv 0$.
%The last term in Eq. (\ref{18total}) is  of the order of the D-wave weight in
%the normalization of the deuteron wave function
% and is neglected here as compared to the S-wave contribution.

 {\it For the $g$-term}  we have got
\begin{eqnarray}
U=0,\, {\bf V}_p=0,\, {\bf V}_n=0,\, W_{ii}=0, \\ \nonumber
 W_{ij}\{S_i,S_j\}= \frac{C^\prime_ng_p+C^\prime_pg_n}{\Pi}
\biggl
\{ ({\hat {\bf n}}_2\cdot{\bf S}) ([{\bf n}_1\times {\bfg\sigma}]\cdot{\bf S}) +
([{\bf n}_1\times {\bfg\sigma}]\cdot{\bf S})({\hat {\bf n}}_2\cdot{\bf S})\biggr \}
\label{g}
\end{eqnarray}
 Furthermore, using the proton spin matrix element
\begin{eqnarray}
<\mu^\prime=\frac{1}{2}|([{\bf n}_1\times {\bfg\sigma}]\cdot{\bf S}|\mu=-\frac{1}{2}>=
%(q'_y\sigma_y+q'_x\sigma_x)S_z\frac{k}{|{\bf n}_1|},
(-iq'_y+q'_x)S_z{k},
\label{gproton}
\end{eqnarray}
we find in the S-wave approximation
\begin{eqnarray}
<\mu'=\frac{1}{2},\lambda'=0|M({\bf q}=0,{\bf Q}; {\bf S}, {\bfg \sigma})|
\mu=-\frac{1}{2},\lambda=1>=
%+<\lambda^\prime=0| W_{ij}\{S_i,S_j\}|\lambda=+1>=
\frac{ik}{\sqrt{2}} \frac{(C^\prime_n g_p+C^\prime_p g_n)q'}{\Pi}S_0^{(0)}(q').
\label{gw}
\end{eqnarray}

%As for the $h$-term, the  D-wave does zero contribution to the interference  with
% the S-wave, and the pure D-wave term is  neglected here.
 The operator (\ref{operator}) does not  depend on the deuteron state.
Therefore, the contribution  of the D-component does not change  the factor
 $C^\prime_n h_p+C^\prime_ph_n$ in Eq. (\ref{hmed})
 and the factor $C^\prime_n g_p+C^\prime_p g_n$ in Eq. (\ref{gw}).
The D-component can contribute  due to the last two terms in Eq.(\ref{18total})  with  $W_{ij}$
 caused by interference  with the S-wave
% that corresponds
%to the  terms with  $S_2^{(1)}$   and 
and  by the  $w^2$-term.
According to analysis performed in Ref.\cite{PK}, at energies $\sim 100$ MeV
  the D-wave contribution  as  well as the
spin-dependent  pN- scattering amplitudes
  are less important than the S-wave contribution and spin-independent pN amplitudes.
% Therefore, in this study we skip cumbersome terms with the D-wave, but will consider them
% in the forthcoming paper.
 Therefore, we postpone  investigation of the D-wave contribution
% which appears in
% the last two terms in Eq.(\ref{18total}) 
%connected to $W_{ij}$ 
   to the next paper.

\subsection{$\widetilde g$ amplitude}
  Thus, the double scattering mechanism with TVPC interaction from Eq. (\ref{TVNN})
leads to the following result for the ${\widetilde g}$ amplitude in the S-wave approximaion
% \begin{eqnarray}
%\label{g5}
%{\widetilde g}=\frac{ik_{pd}}{2\sqrt{\pi}}\frac{1}{4m_p\sqrt{\pi}k_{pN}}
%\int_0^\infty dq q^2S_0^{(0)}(q)[C^\prime_n(q)(g_p-h_p)+C^\prime_p(q)(g_n-h_n)].
% \end{eqnarray}
 \begin{eqnarray}
\label{g5}
{\widetilde g}=\frac{i}{4 m_p\pi}
\int_0^\infty dq q^2S_0^{(0)}(q)[C^\prime_n(q)(g_p-h_p)+C^\prime_p(q)(g_n-h_n)].
 \end{eqnarray}

\section{Numerical results and discussion}
\begin{figure}{}
\includegraphics[width=0.70\textwidth]{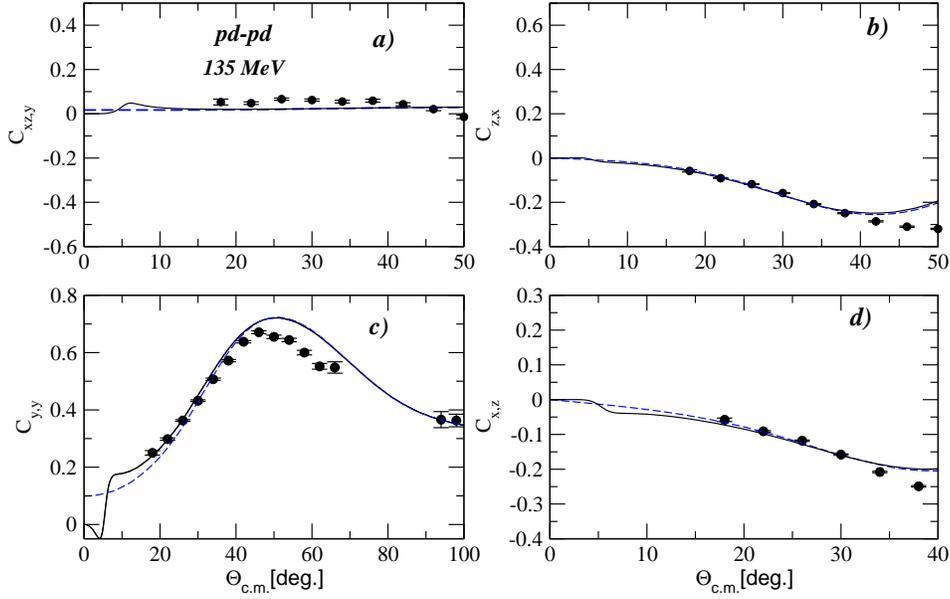}
\caption{(Color online)
Results of our calculation \cite{TUZyaf} of the spin observables $C_{xz,y}$ (a), $C_{z,x}$ (b), $C_{y,y}$ (c)
 and $C_{x,z}$ (d) for the $pd$ elastic scattering
in comparison with the data \cite{przewoski} at 135 MeV: without  (dashed line) and with the Coulomb
interaction included (full).}
\label{fig11}
\end{figure}

 The main aim of this study is to analyze the null-test signal $\widetilde \sigma$ within
 the Glauber theory.
 In order to demonstrate
capability of the Glauber model at energies of the planned COSY
 experiment \cite{TRIC} we
calculated  several spin observables
 of the $pd$ scattering at 135 MeV in comparison with the existing data.
 The results of our calculations for the unpolarized differential cross section,
 vector $A_y$ and tensor $A_{ij}$
 analyzing powers,  and spin correlations parameters $C_{i,j}$, $C_{ij,k}$ given
 in Eqs. (\ref{cijk})
are in reasonable agreement with the available
 experimental data and/or Faddeev calculations
\cite{sekiguchi,przewoski} at 135 MeV and 250 MeV  in the  forward hemisphere
($\theta_{cm}<30 ^\circ$).
Some  of these calculations are shown in Fig. \ref{fig11}.
 We also found  that Coulomb effects taken into account as explained above
 improve agreement with the data on the non-polarized
 differential cross section and vector analyzing powers $A_y^p$ and $A_y^d$
 at these energies at small  angles  $\theta_{cm}\le 20^\circ - 30^\circ$.
 The obtained results  lead us to the conclusion that the Glauber  theory is quite suitable
 for studying  the null-test signal for TVPC effects in the $pd$-scattering  because
 the corresponding signal is not affected by the strong  background of T-even P-even
 interactions.

%%%CHECK 
  The previous study of the null-test signal in $pd$ scattering was performed in Ref. \cite{beyer}.
 The integrated  cross section  $\widetilde \sigma$ was calculated within an
 approach  accounting separately for the elastic channel and the deuteron breakup $pd\to pnp$
 estimated within the single scattering approximation. 
 The double  scattering mechanism was not considered.  
 Extension of the calculation in Ref. \cite{beyer} to higher energies
 above the pion threshold is questionable because the mesons production is not taken into account
 in Ref. \cite{beyer}. On the contrary,
 our approach based on the optical theorem is more general and allows one to
 overcome these drawbacks. In particular, we find that the  $\widetilde \sigma$ observable is
 determined  by the  double scattering mechanism.
% in the  $\widetilde \sigma$ observable. 
% of Ref.\cite{beyer}.
%%%%%%%%% NE vkluycheno
%%% In Ref. \cite{beyer} the contribution of the  elastic channel was found negligible as compared to the breakup one. 
%%%%%%%%%%%%%%%%%%%%%%% 

 Our main result obtained within the Glauber theory is formulated  by
 Eq. (\ref{g5}) for the null-test signal.
It is worth noting that in our approach only the   amplitude $C_N^\prime$
appears  in Eq.(\ref{g5}). Some other T-even P-even $pN$ amplitudes,
which were found in
Ref. \cite{beyer}  to contribute to the TVPC  null-test signal, are absent in
Eq. (\ref{g5}).
 There are two other points worth mentioning in relation  to Eq. (\ref{g5}). First,
 the $g^\prime$-term makes a  zero contribution to ${\widetilde g}$
  and this result is true in general case when both  the S- and D-components of the deuteron
  wave function are taken into account.
 Therefore, an exchange by the
 lightest  meson, that is the $\rho$-meson,
 allowed in general case in the TVPC NN interaction
 \cite{simonius97}
 and, as expected,  makes the most important
 contribution to the TVPC  NN interaction,
 does not contribute
% to the forward $pd$ elastic scattering and
to the null-test signal ${\widetilde\sigma}$.
% too.
 Contribution of other  heavier mesons is usually expected  to be less important due to
 the NN repulsive core  at short distances between nucleons.
 A microscopic T-violating optical
 potential for the nucleon-nucleus interaction
 was derived in Ref. \cite{Engel94} starting from the T-violating $\rho$-meson interaction
  between nucleons. This potential and
 the corresponding coupling constant of the
 $\rho-$meson to the nucleon $\bar g_\rho$ is widely  used \cite{simonius97,huffman}
 as a measure of intensity of the TVPC effects.
 However, as we have shown, for the nucleon-deuteron scattering
 this parameter cannot be applied strightforwardly
 as a scale of the TVPC interactions.

  Strong suppression of the contribution of the $\rho$-meson as compared to the axial $h_1$
 meson was found numerically in the Faddeev calculations \cite{lazauskas} of the null-test signal
 for the $nd$ scattering at 100 keV, but no
% qualitative 
explanation of this result was offered. We suppose that the cause for this suppression
 is the same
 spin-isospin  structure of the scattering amplitude  which leads to the vanishing $\rho$-meson
 contribution in the Glauber approach.
%%%
 A qualitative explanation of the vanishing contribution of the TVPC charge-exchange amplitude 
for the TCPC terms $C_N^\prime$
 is the following. For the strong (TCPC) interaction the charge-exchange  amplitude $M_{pn\to np}({\bf q})$
 appearing in Fig.\ref{fig1},a
  coincides with the charge-exchange  amplitude $M_{np\to pn}({\bf q})$ in Fig. \ref{fig1},b.
 On the contrary, for the
 TVPC interaction caused by $g^\prime$ term the corresponding charge-exchange amplitudes
  have the opposite signs  due to Eqs. (\ref{isospin}). The corresponding deuteron vertices are 
 the same in Fig. \ref{fig1},a and b. 
Taking into account the symmetry in respect of the substitution 
${\bf q}_1\leftrightarrow {\bf q}_2$
 under the sign  of the integral over ${\bf q}'$ 
% and the identical spin-dependence of the strong NN scattering amplitudes with the TCPC ($C_N^\prime$)
% and TVPC ($g^\prime$) terms,
and keeping in mind that spin-dependence of the strong NN scattering amplitudes with 
 the $C_N^\prime$ term
 is identical to that for  the TVPC $g^\prime$ term,
 we find that
the double scattering amplitude in Fig.\ref{fig1},a  differs from that in
   Fig.\ref{fig1},b only by the sign. Therefore, the sum of these diagrams for the $C_N^\prime$
 and $g^\prime$ terms is zero 
\footnote{The TVPC nucleon-nucleon
 interaction with the $g'$ term
 can give  a non-zero contribution to the $pd$ and $nd$ scattering if
 this interaction is included into the
 deuteron wave function. This   is evident from the Fig.\ref{fig1} 
 if the TVPC effects are included only into
  the first (second) deuteron vertex in the both diagrams a and b.
 This dynamics  will be investigated in a special paper.}.

%%%%%%%%%%%%%%%%%%%%%%

 The second point is connected with the role  of  the Coulomb interaction
 in the cross section $\widetilde \sigma$.
Being  a T-even P-even interaction the Coulomb $pp$ scattering
 cannot generate the TVPC amplitude $\widetilde g$ within
the single scattering mechanism, therefore its contribution to $\widetilde g$ is zero
in this approximation.
 In order to include  the Coulomb interaction within the
double scattering mechanism of the Glauber theory one should replace the pure hadronic
 T-even P-even $pp$ amplitude $M_p$  given in Eq.(\ref{pnamp}) by the sum
$M_p+{\tilde f}_{pp}^C$, where ${\tilde f}_{pp}^C$ is
 the properly normalized Coulomb
$pp$-scattering amplitude
(\ref{ppantisym}).
It is evident from the
 spin structure of this amplitudes that the Coulomb term is added
to the
spin-independent  term $A_p$, $A_p\to A_p+{\tilde f}_{pp}^C$ and double spin
terms $B_p$, $G_p$, $H_p$ but does not enter  into
the single spin terms $C_p$ and $C^\prime_p$. However, all  amplitudes
$A_p$, $B_p$, $G_p$, $H_p$ are excluded
from the above derived  formula for the TVPC amplitude (\ref{g5}) due to  the specific spin
 structure of Eq. (\ref{fabtvpc}). As was noted in Sect. \ref{ghterms} the latter statement is also
  true  if  the D-wave  of the deuteron is taken into account.
%(The $h_p$ and $g_p$ terms in Eq. (\ref{g5}),
% as any  TVPC interactions, could not contain the Coulomb
%amplitude additively, because this amplitude  is T-even P-even one).
%The Coulomb pp-amplitude can  be included additively  into  the TVPC $pp$-scattering
% amplitude,
% could be added
%
The only factor in Eq.(\ref{g5})  which contains the Coulomb effects
is  $C_p^\prime(q)$. This is  because
 the Coulomb scattering $pp$ amplitude enters into   the spin-independent term $A_p$, which,
 in  turn,
 enters  into the  amplitude $C_p^\prime$ with the relativistic correction
factor $q/2m_p$  \cite{sorensen} ($m_p$ is the nucleon mass)
 $C_p^\prime=C_p+iA_p\frac{q}{2m_p}$. When substituting this amplitude $C_p^\prime$
into Eq. (\ref{g5}) and making integration over the transferred momentum $q$, one can see that
 due to the presence of the factor $\sim q^3$ in the integrand
 the singularity of the  Coulomb amplitude (\ref{kulonpp}) at $\theta_{pp}=0$ does
 not lead to divergence of the ${\widetilde g}$
 amplitude.

\begin{figure}[ht]
\vspace{1cm}
\includegraphics[width=0.55\textwidth,angle=0]{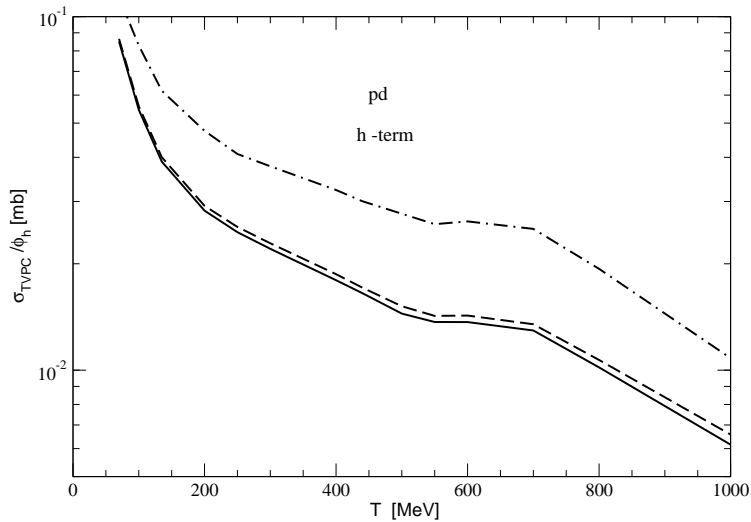}
\caption{The calculated energy dependence of the TVPC cross section $|\widetilde\sigma|$
for the $h$-term in units of the constant $\phi_h=\bar G_h/G_h$ for $m_h=1.17$ GeV,
$\Lambda=2$ GeV in  $h_N$ by Eq. (\ref{hform}) with the
 Coulomb interaction included (dashed line) and excluded (full). The dot-dashed curve is obtained
 at $\Lambda\to \infty$, $m_h^2+{\bf q}^2\to m_h^2$ without account for the  Coulomb effects.
 % and  neglected ${\bf q}^2$-dependence
% of $h_N$ without account for Coulomb effects.
}
\label{difpalpha}
\end{figure}

Energy dependence of $\widetilde \sigma$ is calculated here for the $h$ term in units of the unknown
 $hNN$ coupling constant $\phi_h={\bar G}_h/G_h$  with
$G^2_h/4\pi=1.56$ and  $m_h=1.17$ GeV \cite{lazauskas}.
For the monopole form factor $F_{hNN}$ we used $\Lambda=2$ GeV \cite{beyer}.
The $pN$ scattering amplitudes $C_N$ and $A_N$ were calculated  using the SAID database
\cite{said}.
 The results obtained with and without
allowance for Coulomb effects are shown
in Fig. \ref{difpalpha}. One can see that   the magnitude of
$\widetilde \sigma$  smoothly decreases with increasing beam energy. This is caused be the energy dependence of the
 strong T-even P-even $pN$-scattering  amplitude.  The role of the Coulomb interaction in the null-test
 signal  $\widetilde \sigma$ is  rather unimportant in the considered interval of  energies
 (Fig.\ref{difpalpha}).
 At a beam energy 135 MeV  we obtain 
% $\widetilde\sigma=0.026 \phi_h$ mb
$\widetilde\sigma=0.039 \phi_h$ mb
 and   $\sigma_0=78$ mb.
 Assuming  the accuracy of the experiment allows to measure  the ratio $\widetilde\sigma/\sigma_0$
 at the level $\sim10^{-6}$,
 the bound on $\phi_h$ can be achieved as  $\phi_h\leq0.002$.
 If we neglect some details of the phenomenological  $h$-exchange and  put
 $F_{hNN}=1$ and $(m_h^2+{\bf q}^2)^{-1}\to m_h^2$ in Eq. (\ref{hform}),   the energy dependence of
$\widetilde\sigma$ corresponds to the dot-dashed curve in Fig. \ref{difpalpha}.
 In this case we obtain at 135 MeV
 $\widetilde\sigma=0.063 \phi_h$ mb and, therefore, at the same accuracy of the
 experiement  the bound  is $\phi_h\leq0.0012$.
The $g$-term leads to a very similar energy dependence.
%
%CHECK
 Fig. \ref{difpalpha} shows that the TVPC cross section $\widetilde\sigma$ increases  with
 decreasing energy.	
 Perhaps, at  energies below$\sim  50$ MeV the sensitivity of the experiment \cite{TRIC}
 to the TVPC effects is higher. However, at these low energies the  applicability
 of the Glauber theory to the spin observables of the pd  scattering is not validated,
 and therefore  rigorous calculations like the Faddeev ones are needed.

 Let us consider possible false-effects in the planned experiment \cite{TRIC}.
  One  source of these effects is
  connected with  the non-zero vector polarization of the deuteron
 $p^d_y\not =0$ directed
 along the vector polarization of the proton  beam $p_y^p$. In this case the
 term
 $\sigma_1p_y^p\,p_y^d$  in Eq. (\ref{totalspin}) contributes  to the   asymmetry
corresponding to the cases of $p_y^p P_{xz}^d>0$ and  $p_y^p P_{xz}^d<0$ which is planned  to
 be measured
 in the TRIC experiment \cite{TRIC}.
 According to our calculation, at a beam energy of 135 MeV the total cross sections are  $\sigma_0^t=78.5$ mb,
 $\sigma_1^t=3.7$ mb, $\sigma_2^t=17.4$ mb, and $\sigma_3^t=-1.1$ mb.
 Therefore, the ratio $r=\sigma_1^t/\sigma_0^t$ is  $\approx 0.05$.
 If the TRIC project is going to measure the ratio
 $R_T={\widetilde \sigma}/\sigma_0$
 with an uncertainty about $\leq 10^{-6}$ (upper limit for $R_T$),  one can
 find from  the obtained ratio $r$ that
 the vector polarization of the deuteron
 $p_y^d$  has to be less than $\approx 2\times 10^{-6}$.
 When making this estimation,  we assume that the  background-to-signal ratio  is
 $p_y^d\,\sigma_1^t/{\widetilde \sigma}\sim 10^{-1}$.
The total hadronic polarized
 cross sections $\sigma_i$ ($i=0,1,2,3$) are calculated here using the optical
 theorem.  The Coulomb
 effects for these observables
 can be  taken into account in the line of Ref. \cite{uzjh2009} using the beam
 acceptance angle.

\section{Conclusion}
 Using   the representation of Ref. \cite{rekalo} for the forward elastic $pd$ scattering
 amplitude  and including
 the  phenomenological TVPC-term in the most general form, we show
 on the basis of the optical theorem
 that this term generates  an extra spin dependent total cross section
 of the $pd$ scattering.
 % $\widetilde \sigma$ in the total $pd$ cross section.
  This additional term  is  zero if the
 TVPC interaction is absent and  non-zero only in the
 presence of the TVPC interaction. Earlier this conclusion was found in different
 representations \cite{conzett,beyer}.
 Obviously,  this null-test signal for T-invariance violation is not affected by the
 initial or/and final state interactions, because it is derived from a genuine $pd$-scattering
 amplitude considered  beyond  the perturbation theory. Furthermore, using the Glauber
 theory  we show that  (i) the TVPC
 %T-violating P-parity conserving
  interaction caused by
 the $\rho$-meson exchange
 does not contribute to the null-test observable $\widetilde \sigma$ and (ii)
 the Coulomb interaction   does not lead to divergence of this observable. Numerical
 calculation of the energy dependence of $\widetilde \sigma$ shows that the choice  of $\sim 100$ MeV
 made in Ref. \cite{TRIC} is more preferable than $\sim 1$ GeV.  
 If the cross section $\widetilde \sigma$ will
  be measured   with the planned  accuracy \cite{TRIC},
% $\widetilde \sigma/\sigma_0\sim 10^{-6}$ then 
the bounds on T-odd coupling constants of the NN-interaction can be achieved as
  $\phi_h\leq (1\div 2)\times 10^{-3}$.

{\bf Acknowledgments}. We are thankful to C.~Wilkin for reading this paper and making
 helpful remarks, V.~Gudkov for comments on Eq. (\ref{VNNh}), I.~Strakovsky for consultation
 concerning the SAID database, and
 P.D.~Eversheim and Yu.~Valdau for  interest in this work.

\end{document}